\documentclass[aps,prl,reprint,groupedaddress,superscriptaddress,showpacs]{revtex4-1}

\usepackage{graphicx}
\usepackage{amssymb}
\usepackage{amsmath}
\usepackage{color}
\usepackage{ulem}	%For double underlined symbols

\begin{document}

\title{
Suppression of bacterial rheotaxis in wavy channels
}

\author{Winfried Schmidt}
\affiliation{Theoretische Physik, Universit\"at Bayreuth, 95440 Bayreuth, Germany }
\affiliation{Universit\'e Grenoble Alpes, CNRS, LIPhy, F-38000 Grenoble, France}

\author{Igor S.\ Aranson}
\affiliation{
	Departments of Biomedical Engineering, Chemistry, and Mathematics, Pennsylvania State University,
	University Park, PA 16802, USA
}

\author{Walter Zimmermann}
\affiliation{Theoretische Physik, Universit\"at Bayreuth, 95440 Bayreuth, Germany }

%\date{\today}

\begin{abstract}	
Controlling the swimming behavior of bacteria is crucial, for example, to prevent contamination of ducts and catheters.
We show the bacteria modeled by deformable microswimmers can accumulate in flows through straight microchannels either in their center or on previously unknown attractors near the channel walls.
We predict  a novel resonance effect for semiflexible microswimmers in flows through wavy microchannels.	 
As a result, microswimmers can be deflected in a controlled manner so that they swim in  modulated channels distributed over the channel cross-section rather than localized near the wall or the channel center. 
 Thus, depending on the flow amplitude, both upstream orientation of swimmers and their accumulation at the boundaries which can lead to surface rheotaxis are suppressed.
Our results suggest new strategies for controlling the behavior of live and synthetic swimmers in microchannels.
\end{abstract}


\maketitle

Bacteria are among the most wide-spread microorganisms in nature. 
One of the remarkable properties of motile bacteria is the ability to reorient their bodies against the flow and   swim upstream,  i.e. positive rheotaxis \cite{rusconi2014bacterial, mino2018coli, Junot_2019}. This often detrimental behavior leads to contamination of ducts and catheters that may lead to bacterial infections \cite{mathijssen2019oscillatory, Figueroa-Moraleseaay0155}.   Positive rheotaxis  occurs for sperm cells and plays an important role in the reproduction process \cite{doi:10.1098/rspb.1961.0014,MIKI2013443,kantsler2014rheotaxis,bukatin2015bimodal,waisbord2021fluidic}. Recently, rheotaxis-like behavior was observed for synthetic self-propelled particles, although the mechanisms are not necessarily similar to that of bacteria and sperm cells \cite{Palaccie1400214,ren2017rheotaxis,uspal2015rheotaxis,brosseau2019relating,baker2019fight,rubioself}. Despite the  importance of rheotaxis  for human and animal health and reproduction, many underlying mechanisms are not clear.

The familiar dynamics of a rigid microswimmer in a planar Poiseuille flow is determined
by the interplay between  the swimmer's speed
and the flow vorticity  \cite{PhysRevLett.108.218104, Zoettl2013,UPPALURI20121162, Junot_2019}. 
Two different types of motion have been identified: (i) 
The  swinging motion is characterized by sinusoidal swimmer trajectories around the channel center.
It occurs  for  low flow  strengths  compared to the swimming speed. (ii) 
 The tumbling motion is observed for large  flow 
velocities where the flow vorticity is sufficiently strong
to reorient the swimmer before it reaches the channel center, resulting in complete rotations of the swimmer. 
 
\begin{figure}[htb]
\begin{center}
	\includegraphics[width=0.98\columnwidth]{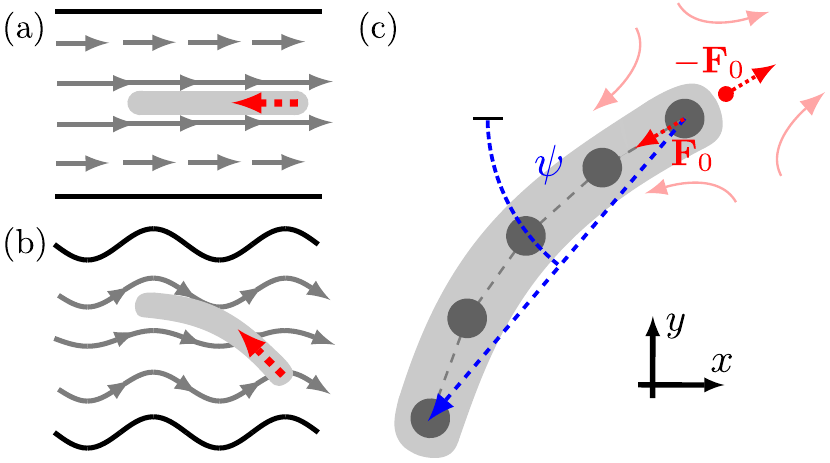} \vspace{-.5cm}
\end{center}
\caption{ 
	(a) A flexible  swimmer (light gray)   swims against (red dashed arrow) straight flow lines at the channel center  (channel walls  black lines,  swimmer size enhanced). (b) A wavy channel  suppresses rheotaxis. 
	(c) Deformable  swimmer (light gray), composed of $N=5$ beads (dark gray)  in the $x$-$y$-plane, with its mean orientation  angle $\psi$ (blue dashed line). The force dipole at its rear (pair of red arrows) creates a pusher-type flow field (light red dotted arrows) and thrust along the instantaneous swimming direction, given by ${\bf F}_0$.
}
\label{fig_sketch}
\end{figure}

Many microswimmers are deformable. They, for instance, bend their bodies  \cite{Wang9182, 10.1371/journal.pone.0083775} for the purpose of self-propulsion, as in the case of \textit{Spiroplasma} \cite{PhysRevLett.99.108102}, or  have flexible flagella \cite{tournus2015flexibility, C9SM00717B,potomkin2017flagella}. Flexible elongated microswimmers migrate 
transversely to streamlines in plane Poiseuille flow  \cite{tournus2015flexibility, C9SM00717B}, similar to semiflexible polymers \cite{farutin2016dynamics, slowicka2013lateral},  vesicles \cite{PhysRevE.77.021903}, droplets \cite{Leal:1980.1}   or capsules in oscillating shear flows \cite{Laumann:2017.1}. 
 Semiflexible microswimmers are able to migrate across streamlines toward the channel center where they reorient against the flow \cite{tournus2015flexibility, C9SM00717B}, as shown in Fig.\ \ref{fig_sketch}(a). This type of rheotaxis can result in  swimmer accumulation at the channel center.
 We find two  additional attractors for semiflexible microswimmers, located near  the plane channel walls.
The position of the repeller separating inward and outward swimming directions  depends on the flow velocity.
In wavy Poiseuille flows [outlined in Fig.\ \ref{fig_sketch}(b)] we find 
 a novel resonance for specific flow velocities and channel geometries, resulting in depletion of swinging swimmers from the channel center.
Moreover, by the wavy-induced tumbling motion of swimmers, migration to the peripheral
attractors is suppressed in a controlled manner.

 A semiflexible microswimmer in a Newtonian fluid with viscosity $\eta$  is modeled by $N$ small spheres  with radius $a$  at positions ${\bf r}_i = (x_i, y_i, z_i)$ ($i = 1,...,N$) [cf.\ Fig.\ \ref{fig_sketch}(c)], and its center at ${\bf r}_\text{c} = \sum_{i=1}^{N} {\bf r}_i/N$.
 The undeformed swimmer is straight with length $L_0 = (N-1)b + 2a$, with equilibrium distance of two neighboring beads  $b$.
 The translational  $\dot{{\bf r}}_i$ and angular velocities ${\boldsymbol \Omega}_i$ of each bead are 
\begin{align}
\notag \dot{{\bf r}}_i = {\bf u} ({\bf r}_i) &+ \sum_{j=1}^{N+1} {\boldsymbol \mu}_{ij}^\text{tt} {\bf F}_j + \sum_{j=1}^{N} {\boldsymbol \mu}_{ij}^\text{tr} {\bf T}_j, \\
\label{eq_motion}{\boldsymbol \Omega}_i = \frac{1}{2} \nabla \times {\bf u} ({\bf r}_i) &+ \sum_{j=1}^{N+1} {\boldsymbol \mu}_{ij}^\text{rt} {\bf F}_j + \sum_{j=1}^{N} {\boldsymbol \mu}_{ij}^\text{rr} {\bf T}_j
\end{align}
with the mobility matrices ${\boldsymbol \mu}_{ij}$ \cite{wajnryb_mizerski_zuk_szymczak_2013, zuk_wajnryb_mizerski_szymczak_2014} and bead torques ${\bf T}_i$  coupling bead rotations to the swimmer configuration,  as given in Ref.\,\cite{supplement}. ${\bf u} ({\bf r}_i)$ is the background flow as described below. 
The bead forces are  ${\bf F}_i = - \nabla_i \left( V_i^\text{H} + V_i^\text{B} \right)$ with the 
harmonic spring potential 
 $V_i^\text{H}$ and spring constant $k$ between neighboring beads. $V_i^\text{B} = - \kappa/2 \ln \left( 1 - \cos \alpha_i \right)$ is a bending potential \cite{PhysRevE.51.2658}  with 
bending rigidity $\kappa$  and opening angle  $\alpha_i$ of the chain at sphere $i$. We assume an inextensible swimmer with large $k$  and $\kappa$ smaller to allow for bending.
Self-propulsion is implemented by a force ${\bf F}_0 = F_0 \hat{\bf{e}}_s$  acting on the $N$-th bead in the chain
with  unit vector in swimming direction $\hat{\bf{e}}_s := \left({\bf r}_{N-1} - {\bf r}_{N}\right) / |{\bf r}_{N-1} - {\bf r}_{N}|$. 
This driving force ${\bf F}_0$  is balanced for a force free swimmer by  its antiparallel counterpart, $-{\bf F}_0$, acting on 
the  counter-force point at ${\bf r}_N - 2 a \hat{\bf{e}}_s$. Depending on the sign of $F_0$, this pair of forces pushes the bacterial body  in front of it and creates the flow field of a pusher ($F_0 > 0$), or drags the body behind it and creates a puller-type flow field ($F_0 < 0$) \cite{Elgeti_2015, doi:10.1063/1.4944962}.  The flow disturbance caused by the counter-force point
 accounts for the $N + 1-$st contribution to the translational and angular velocities in Eq.\ (\ref{eq_motion}).
The     swimming speed $v_0$ depends linearly on $F_0$  \cite{supplement}.
For simplicity, we restrict our analysis to the $x$-$y$ plane  and  on  pushers, as for rod-shaped swimmers like the bacteria \textit{E. coli} or \textit{B. subtilis}. The swimmer dynamics is characterized by the angle $\psi$ between its end-to-end vector and the $x$-axis. Pushers and pullers behave similar for 
large parameter ranges.

 We consider a serpentine channel geometry with the same modulation wavelength $\lambda$ and phase  for the two opposite walls   [cf.\ Fig.\ \ref{fig_sketch}(b)] at
\begin{equation}\label{eq_serpentine_walls}
y_\text{w} (x) = d \left[\pm 1 + \varepsilon \sin \left( \frac{2 \pi x}{\lambda} \right) \right].
\end{equation}
 Here, $d$ is the channel half-height and $\varepsilon$ the dimensionless  modulation amplitude. That is in contrast to channels in Ref.\ \cite{PhysRevLett.122.128002} 
 on cross-stream migration (CSM) of capsules and red blood cells, where modulations of the opposite walls are shifted by half a wavelength. 
 For $\varepsilon \ll 1$, we determine the flow field ${\bf u} ({\bf r})=(u_x,u_y,0)$  by a perturbation expansion, 
\begin{eqnarray}\label{eq_wavy_poiseuille}
\notag u_x(x,y) &=& u_0 \left( 1-\frac{y^2}{d^2} + \varepsilon \ U_1(x,y)\right) \\
u_y(x,y) &=& u_0 \varepsilon \ U_2(x,y),
\end{eqnarray}
with flow amplitude $u_0$. The functions  $U_1(x,y)$ and $U_2(x,y)$ are given in   
 \cite{supplement}  and
the parameters in \cite{parameters_incl_plane}.

\begin{figure}[htb]
	\begin{center}
		\includegraphics[width=0.98\columnwidth]{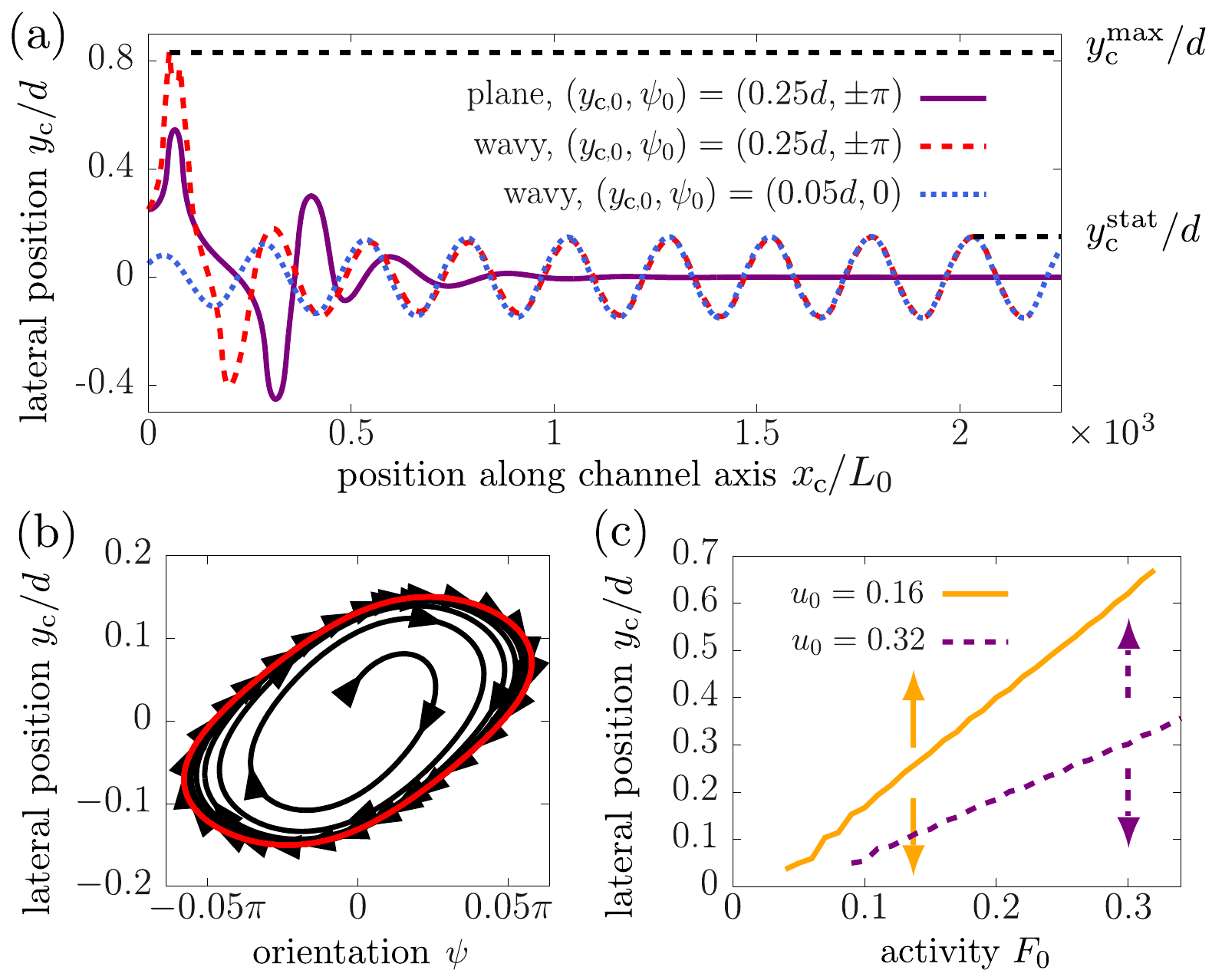} \vspace{-.5cm}
	\end{center}
	\caption{
		(a): A semiflexible microswimmer with $F_0 = 0.6$ and $\kappa = 0.5$ approaches the center  of a plane channel
		along the violet solid trajectory, orienting upstream. In wavy channels  this fixed point is replaced by a limit cycle (red dashed and blue dotted lines). Maximum ($y_\text{c}^\text{max}$) and steady-state amplitude ($y_\text{c}^\text{stat}$) are also shown. (b): Phase space trajectory (black) in wavy flow for initial conditions $(y_{\text{c},0}, \psi_0) = (0.05d, 0)$ and the limit cycle (red).
		(c):  Repeller for tumbling swimmers
		for $u_0 = 0.16$  (orange solid line) and $u_0 = 0.32$  (violet dashed line). Arrows illustrate the migration direction.
}
	\label{fig_wavy1} 
\end{figure}

Fig.\ \ref{fig_wavy1}(a)     compares  trajectories of a semiflexible swimmer in  unbounded
 planar ($\varepsilon=0$) 
and wavy Poiseuille flow with $\varepsilon=0.1$. 
In planar Poiseuille flow it first tumbles, then transitions to swinging, and finally approaches the
 fixed point $(y_\text{c},\psi) = (0,0)$ (upstream orientation at the channel center). 
  In a wavy channel, the non-zero flow vorticity causes the swimmer to oscillate periodically about its mean upstream orientation, resulting in a swinging motion.
Choosing $u_0 > v_0$,  swimmers drift downstream in both planar and wavy  flows.
 The short-time transient in the wavy flow depends on the swimmer's initial position and orientation, but the long-time behavior does not [cf.\ red dashed and blue dotted trajectory in Fig.\ \ref{fig_wavy1}(a)].
In the following, we refer to the  constant long-time swinging amplitude as $y_\text{c}^\text{stat}$ and to the maximum of the oscillation  during the transient regime as $y_\text{c}^\text{max}$, as indicated in Fig. \ref{fig_wavy1}(a).
Fig.\ \ref{fig_wavy1}(b) shows a phase space trajectory for a swimmer starting upstream oriented from a lateral position near the center of the channel that converges to a periodic trajectory (limit cycle).

In planar Poiseuille flow, a rigid swimmer is described by a Hamiltonian system with periodic phase-space orbits that depend on the initial conditions \cite{PhysRevLett.108.218104, zottl2013periodic}.
A semiflexible swimmer  breaks the periodicity of its trajectory  \cite{tournus2015flexibility}. 
This results in the inward drift during tumbling and  the transition to swinging with a decaying amplitude  \cite{supplement}.
 Swimmer reorientation against the flow is thus the result of an interplay of its shear-flow induced deformation and  self-propulsion $F_0$.  
This inward drift stands in contrast to the outward migration of passive soft particles with elongated shapes, such as flexible fibers and elongated vesicles \cite{farutin2016dynamics, slowicka2013lateral}. 
This type of migration originates from the particle's hydrodynamic self-interaction  and its shear-rate-induced deformation.  
These two competing migration mechanisms result in a repeller, as shown in Fig.\ \ref{fig_wavy1}(c) that separates outward and inward-directed trajectories.
Whereas the mechanism of passive CSM dominates for small $F_0$ and large distances from the channel center, the activity-induced inward CSM  outweighs the outward CSM for growing  $F_0$. 
 The repeller for $u_0 = 0.16$ lies above the repeller for $u_0 = 0.32$,  since weaker flows result in a larger relative influence of activity.

\begin{figure}[htb]
	\begin{center}
		\includegraphics[width=0.98\columnwidth]{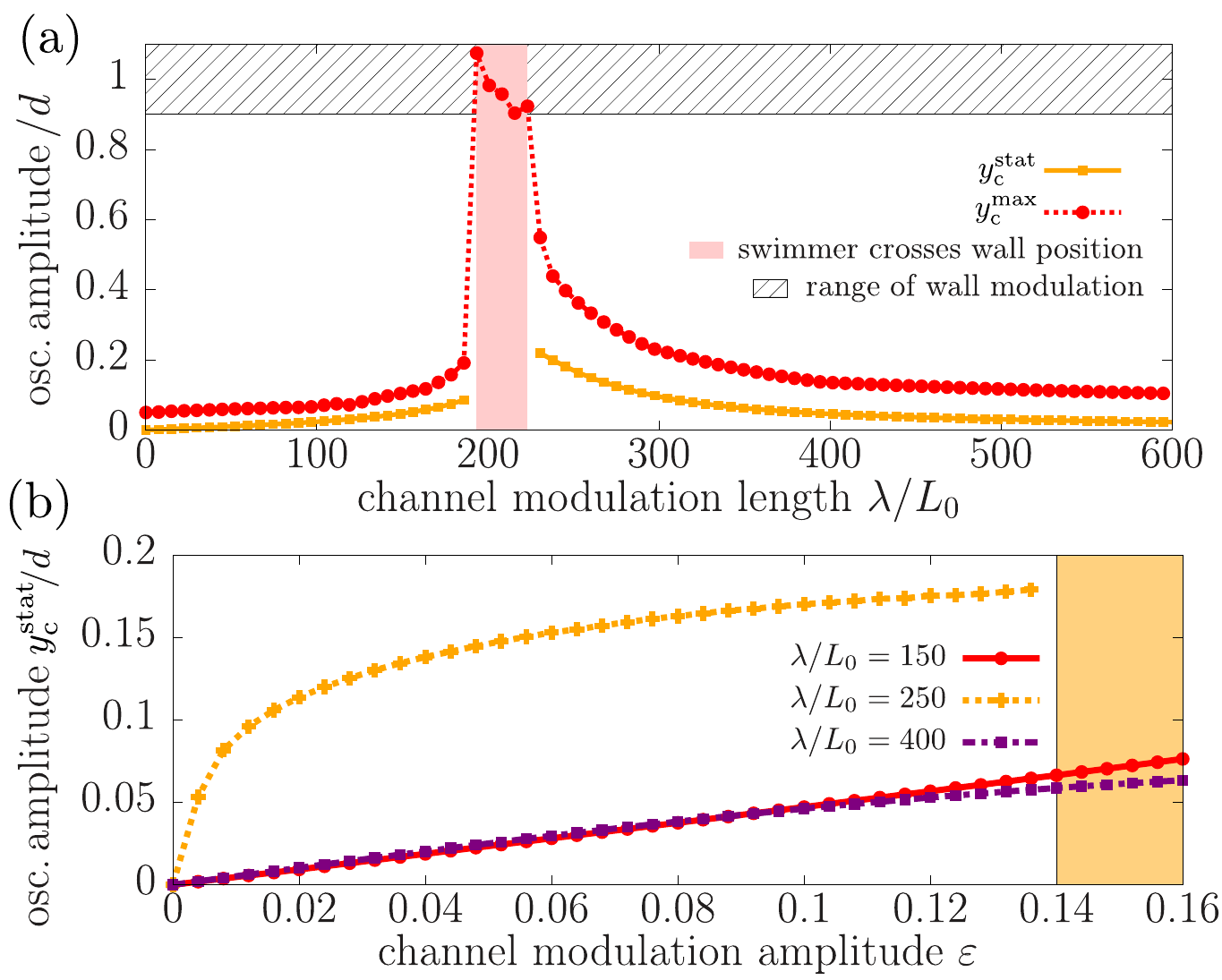} \vspace{-.5cm}
	\end{center}
	\caption{
		(a): Resonance curve with steady-state ($y_\text{c}^\text{stat}$, orange bold line, squares) and maximum  amplitude ($y_\text{c}^\text{max}$, red dotted line, circles) vs  modulation length $\lambda$ in units of the 
		swimmer length $L_0$. Wall  modulation for  $\varepsilon = 0.1$  in the black shaded range.
		(b): $y_\text{c}^\text{stat}$ vs  modulation amplitude $\varepsilon$ for $\lambda / L_0 = 150$ below (red bold lines, circles) 
		and $\lambda / L_0 = 400$ beyond the resonance wavelength (violet dashed-dotted lines, squares).
 For $\lambda / L_0 = 250$ (orange dotted lines)  the swimmer crosses the wall position in the orange range.
	}
	\label{fig_wavy2} 
\end{figure}

We now analyze the swimmer behavior in the wavy flow as a function of  modulation length $\lambda$ and amplitude $\varepsilon$.  We compute
 $y_\text{c}^\text{stat}$ by the maximum of the  magnitude of $y_\text{c}(t)$ beyond the transient regime in $t \in [1, 2] \times 10^6$.
 For small oscillations, the swinging frequency of an elongated microswimmer in planar Poiseuille flow is given by $\omega_0 = \sqrt{u_0 v_0(1-G)} / d$ \cite{zottl2013periodic}, with geometry factor $G = \left( r_\text{p}^2 - 1 \right) / \left( r_\text{p}^2 + 1 \right)$ and swimmer aspect ratio $r_\text{p} = 1 + (N-1) b / (2 a)$.
 An additional frequency $\omega_\text{Ch}$ is imposed on the swimmer when it moves along 
  wavy streamlines. Assuming a perfect upstream swimmer orientation 
   yields  $\omega_\text{Ch} \approx 2 \pi |u_0 - v_0 |/\lambda$. In our system, $\omega_0$ can be interpreted as the oscillator eigenfrequency and $\omega_\text{Ch}$ the frequency of an external periodic drive with amplitude $\varepsilon$. We expect the swinging amplitude to peak in the resonance case of $\omega_\text{Ch} \approx \omega_0$, which 
determines a  resonance wavelength via
\begin{equation}\label{eq_resonance_wavelength}
\lambda_\text{res} \approx \frac{2 \pi d |u_0 - v_0|}{\sqrt{u_0 v_0 \left( 1 - G \right)}}.
\end{equation}
Fig.\ \ref{fig_wavy2}(a) shows the swinging amplitude vs $\lambda$. For small  wavelengths, both the initial and steady-state response of the system are small, with  $y_\text{c}^\text{stat} \rightarrow 0$ for $\lambda \rightarrow 0$. This case corresponds to a very large $\omega_\text{Ch}$.
Increasing $\lambda$ causes increasing  maximum and steady-state  oscillation amplitudes, culminating in a peak of $y_\text{c}^\text{max}$ in the range of $193 \lesssim \lambda / L_0 \lesssim 223$ [light red area in Fig.\ \ref{fig_wavy2}(a)]. For these $\lambda$, the transient swimmer response is large enough that one of its beads reaches the wall position. In this case, we stop the simulation in unbounded flows.  The influence of repulsive swimmer-wall interactions is discussed below.
For further growing $\lambda$, $y_\text{c}^\text{max}$ and $y_\text{c}^\text{stat}$ decrease monotonically and approach a small but finite amplitude for large  wavelengths.

With the parameters listed in \cite{parameters_incl_plane}, we obtain $\lambda_\text{res} \approx 257 L_0$ from Eq.\ (\ref{eq_resonance_wavelength}) which is close to the resonance  region in Fig.\ \ref{fig_wavy2}(a). 
The difference between this theoretical prediction and the numerics arises from the  assumption of perfect upstream  orientation and a   constant swimmer position at the channel center. For increasing swinging amplitudes, i.e., for  $\lambda \approx \lambda_\text{res}$, the swimmer on average visits positions further away from the channel center more often where the flow is slower. 
Thus, $\omega_\text{Ch}$ is effectively smaller than assumed above. 

Increasing $\varepsilon$ results in a growing size of the limit cycle,  as shown in Fig.\ \ref{fig_wavy2}(b). That applies to $\lambda$ smaller, larger, and close to the $\lambda_{\text{res}}$ [cf.\ Fig.\ \ref{fig_wavy2}(a)]. In the latter case, we observe crossings of wall positions of  the swimmer during the initial transient for $\varepsilon > 0.14$. 
 For the effect of activity on the swimmer behavior in the wavy channel see{\tiny } \cite{supplement}.

\begin{figure}[htb]
	\begin{center}
		\includegraphics[width=0.98\columnwidth]{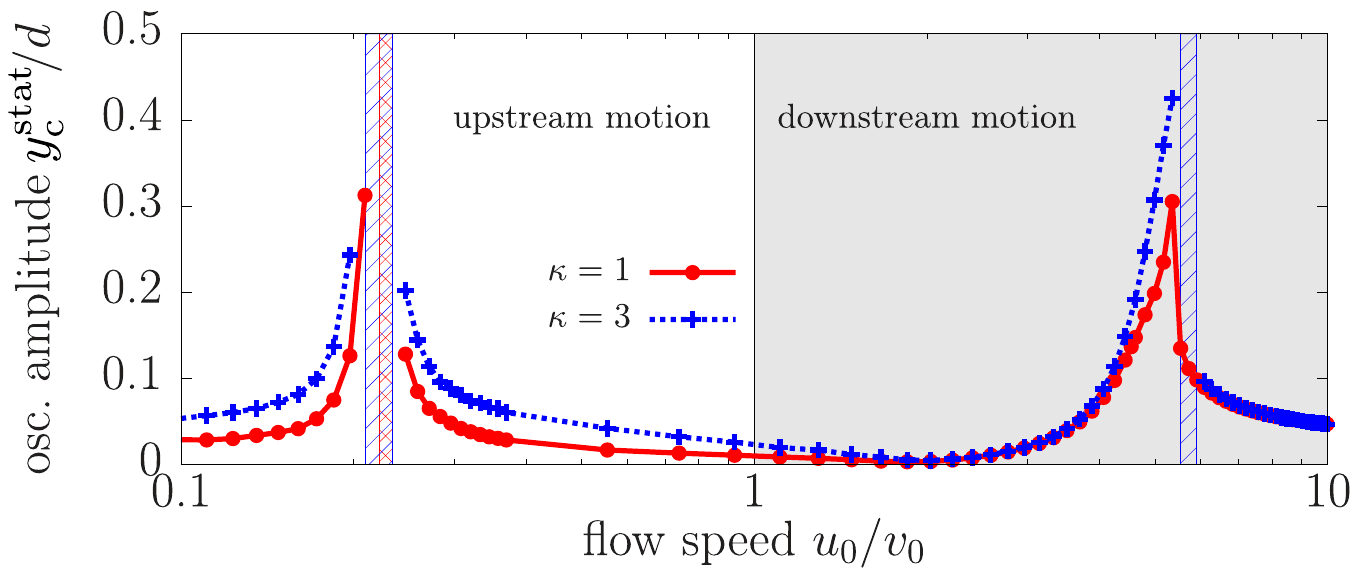} \vspace{-.5cm}
	\end{center}
	\caption{
		  The steady-state swinging amplitude $y_\text{c}^\text{stat}/d$ vs velocity ratio $u_0/v_0$
		  is larger for stiffer swimmers ($\kappa = 3$, blue dotted lines, crosses) than for softer ones ($\kappa = 1$, red bold lines, circles),  with upstream  (white, left panel) and downstream resonance  (gray, right panel).
		 The swimmer exceeds the wall position in certain ranges of $u_0/v_0$
		  (blue dashed boxes for $\kappa = 3$ and red double-dashed box for $\kappa = 1$).
	}
	\label{fig_wavy3} 
\end{figure}

Apart from the channel geometry,  the experimentally controllable flow strength  
significantly impacts the swimmer behavior  in the wavy channel. We choose $\lambda = 150 L_0$, a  wavelength smaller than  $\lambda_{\text{res}}$ in Fig.\ \ref{fig_wavy2}(a), and show in Fig.\ \ref{fig_wavy3}  the steady-state swinging amplitude  vs the  normalized flow amplitude for different  rigidities $\kappa = 1, 3$.
We observe resonant behavior for  both upstream ($u_0 < v_0$) and downstream motion ($u_0 > v_0$). Depending on $\kappa$, the resonant oscillations can become large enough to trigger a crossing of one of the channel wall  positions. Flow speed ranges above and below the respective resonance ratio $u_0 / v_0$ are characterized by a comparably small $y_\text{c}^\text{stat}$. Assuming fixed $\lambda$, we solve Eq.\ (\ref{eq_resonance_wavelength}) for $u_0 / v_0$, yielding
\begin{equation}\label{eq_theo_prediction_u0}
	\frac{u_0}{v_0} = 1 + \frac{\alpha^2}{2} \left( 1 - G \right) \pm \alpha \sqrt{\left( 1 - G \right) + \frac{\alpha^2}{4} \left( 1 - G \right)^2},
\end{equation} 
where $\alpha := \lambda /(2 \pi d) > 0$ depends only on the channel's geometry. Eq.\ (\ref{eq_theo_prediction_u0}) has the two solutions $u_0/v_0 = 4.54$ (downstream drift) and $u_0 / v_0 = 0.22$ (upstream swimming). Both give a good approximation for the location of the numerically obtained maxima in Fig. \ref{fig_wavy3}.  Our  predictions for $\omega_0$ and $\omega_\text{Ch}$ agree well with the numerically obtained Fourier spectra of the swimmer trajectory  \cite{supplement}.

\begin{figure}[htb]
	\begin{center}
		\includegraphics[width=0.98\columnwidth]{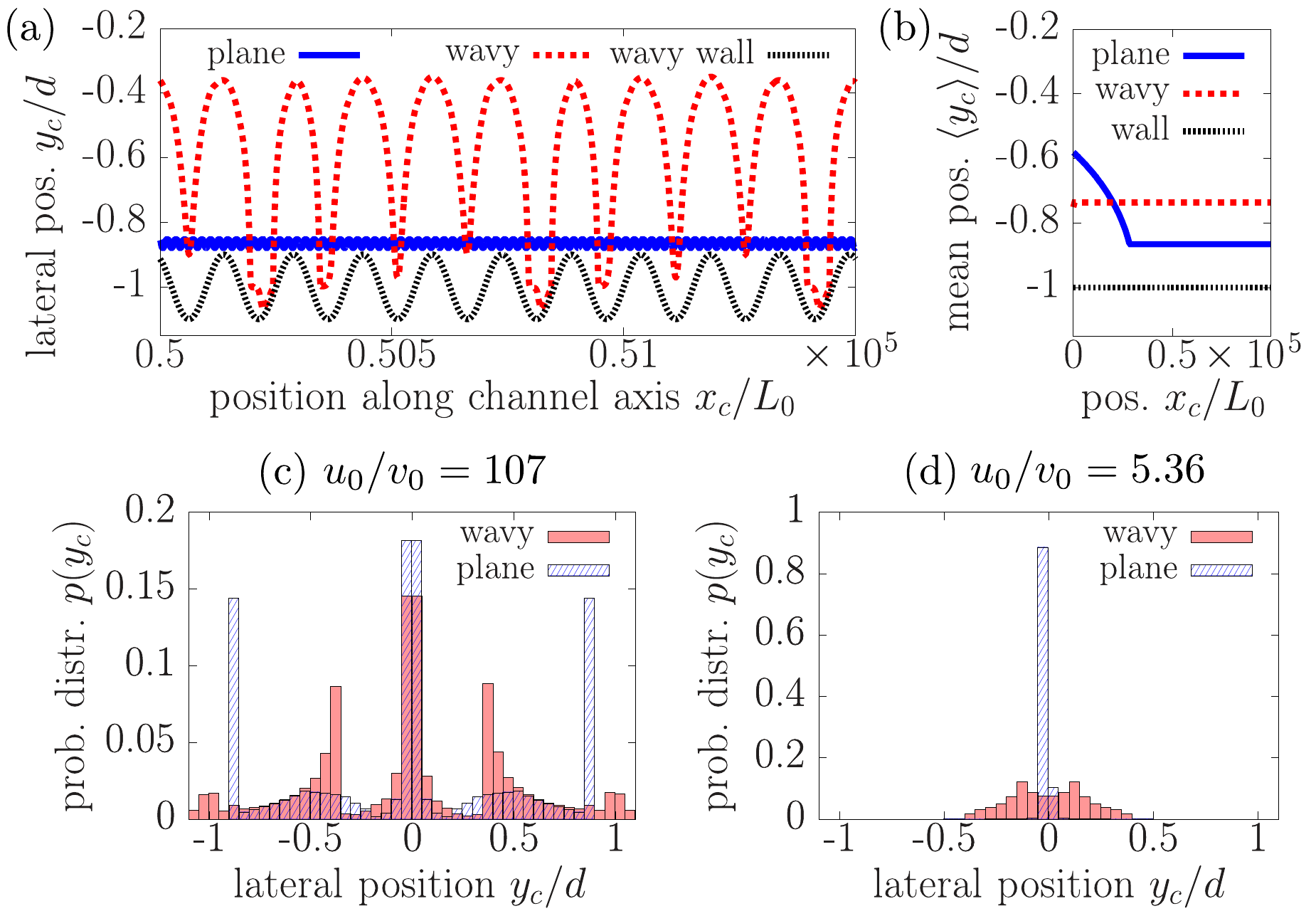} \vspace{-.5cm}
	\end{center}
	\caption{ (a) Individual and (b) mean trajectories for  a swimmer starting at $y_{c,0}=-0.6d$ in straight (blue bold line) and wavy streamlines (red dashed line) with  $u_0 = 107 v_0$. Black dotted line is 
		the wavy channel boundary in (a) and its mean at $y_\text{w}=-1$ in (b). 
		(c) Probability distribution $p(y_c)$ of a swimmer in straight (dashed blue bars)
		 and wavy streamlines (bold red bars) for  $u_0/v_0 = 107$ and in (d) for  $u_0/v_0 = 5.36$.	 
	}
	\label{fig_wavy4} 
\end{figure}

 We now include  a short-range repulsive  wall potential \cite{doi:10.1063/1.1674820} in our simulations, as described in \cite{supplement}. 
For  a large ratio $u_0/v_0 = 107$ we find in plane channels  a repeller at $y \approx \pm d/4$.
Swimmers beyond the repeller  migrate closer to the wall and then tumble around a   constant $y$-position  [blue bold trajectory in Fig.\ \ref{fig_wavy4}(a)]. 
In wavy flows,  tumbling has a much larger amplitude and a periodicity of  $4 \lambda$ [red dashed trajectory in Fig.\ \ref{fig_wavy4}(a)].
Fig.\ \ref{fig_wavy4}(b) shows swimmer trajectories averaged over $4 \lambda$ on a longer time scale. One finds an initial outward drift in  plane flows, whereas no drift occurs in  wavy channels,  where $\langle y_c \rangle$ is closer to the center for large times.
In plane  flows, the swimmer migrates by
a  lateral distance of $0.32d$ while traveling $2.9 \times 10^4 L_0$ in $x$-direction [blue bold trajectory in Fig.\ \ref{fig_wavy4}(b)]. By contrast, lateral motion in wavy flows is faster: The large tumbling amplitude enables the swimmer to move up to $0.71d$ in $y$-direction while being advected downstream only $86 L_0$.

Figs.\ \ref{fig_wavy4}(c) and (d) show the  lateral probability distribution $p (y_\text{c})$  for large $u_0/v_0$ and a $20$ times smaller ratio, $u_0/v_0 = 5.36$. 
We obtain $p (y_\text{c})$ by averaging individual swimmer distributions for $10$ different initial positions $y_{c,0} \in [-d/2, d/2]$.
Each individual distribution is determined during the simulation time $t_\text{end} = 3 \times 10^7$.
 The repeller causes  swimmers in plane channels to accumulate for larger $u_0/v_0$ either at  their center or at  an attractor close to each wall [three peaks in Fig.\ \ref{fig_wavy4}(c)]. The small probabilities between the peaks of $p (y_\text{c})$  result from the transient CSM to the attractors. By contrast, in wavy channels the large tumbling amplitude shown in Fig.\ \ref{fig_wavy4}(a) leads to small values of $p(y_\text{c})$ near the walls even for a large velocity ratio.
For smaller $u_0/v_0$, swimmers migrate to the center of plane channels for all initial positions, resulting in a single peak of $p (y_\text{c})$ at $y_\text{c} = 0$ in Fig.\ \ref{fig_wavy4}(d). This behavior is changed in wavy flows where  the above described swinging motion [see Fig.\ \ref{fig_wavy1}(a)] broadens $p (y_\text{c})$ 
and reduces the  swimmer probability at the channel center significantly.

In this work we analyzed the behavior of elongated semiflexible microswimmers, such as bacteria, 
in flows through both plane and wavy microchannels.
In planar channels, at lower flow velocities, swimmers concentrate at their center while  orienting upstream.
At higher flow velocities, we predict two additional attractors closer to the walls for tumbling swimmers, which coexist with the attractor at the channel center.
The proximity of the attractors to the walls may promote the formation of bacterial biofilms at the boundaries \cite{flemming2016biofilms} as well as upstream migration due to surface rheotaxis \cite{PhysRevLett.98.068101, KAYA20121514}. 
This is suppressed by  wavy boundaries which lead to swimmer depletion close to the walls.  
For certain ranges of the ratio of the  flow strength to swimming speed, we discovered a resonance effect induced by the wavy streamlines. This and the associated swinging/tumbling motion  can be controlled by the flow velocity, the swimmer's speed and size, and the boundary modulation.
Thereby swimmers cross the entire channel periodically and the probability distribution  becomes broad. This can, for example, prevent the formation of swimmer clusters \cite{PhysRevE.90.063019}. In addition, near the resonance, bacteria are forced to hit the walls where they can be killed, e.g., by nanopillars \cite{michalska2018tuning} or antibacterial surface coatings  \cite{vasilev2009antibacterial}. We expect hydrodynamic swimmer-wall interactions \cite{kurzthaler_stone} to affect our results quantitatively,  but not fundamentally.
Possible emergent behavior due to noise effects
\cite{ezhilan_saintillan_2015,PhysRevResearch.2.033275} and chirality of
flagella \cite{mathijssen2019oscillatory} are the subject of future
studies.

 W.S. thanks for support DAAD, W.S.\ and W.Z.\ the French-German University (Grant No.\ CFDA-Q1-14, ``Living fluids'') and the Elite Study Program Biological Physics, and A.\ F{\"o}rtsch and M.\ Laumann for inspiring discussions.  The research of I.S.A. was supported by the NSF awards PHY-2140010.

%\bibliographystyle{prsty}

%\bibliography{bib_swim.bib}

\begin{thebibliography}{10}
\bibitem{parameters_incl_plane}
	{ Simulation parameters for plane flows: Time step $\Delta t=0.01$, maximum
		flow speed $u_0=0.162$, channel half height $d=20$, fluid viscosity $\eta=1$,
		number of beads $N=5$, equilibrium distance of two neighboring beads $b =
		1.095$, bead radius $a = 0.5$ (aspect ratio $r_\text{p}=5.38$, geometry
		factor $G=0.933$), activity $F_0 = 0.6$ (intrinsic swimming speed $v_0 =
		4.872 \times 10^{-3}$), harmonic spring stiffness $k = 10$, bending rigidity
		$\kappa=0.5$, torque strength $\kappa_\text{t} = 10$, initial swimmer
		orientation $\psi_0=\pm \pi$, initial swimmer lateral position $y_0 = 5$. The
		migration direction is determined by the sign of the migration velocity which
		is obtained by a linear fit to $y_\text{c} (t)$ for $t \in [0, 10^5]$.\\
		Simulation parameters in wavy flows: Activity $F_0 = 1$ (intrisic swimming
		speed $v_0 = 0.0162$), bending rigidity $\kappa = 3$, dimensionless
		modulation amplitude $\varepsilon = 0.1$, modulation length $\lambda = 807$,
		initial swimmer orientation $\psi_0=0$, initial swimmer position $(x_0, y_0)
		= (0, 1)$, simulation time for each trajectory $t_\text{end} = 2 \times
		10^6$. Remaining parameters as in simulations in plane flows.}
\end{thebibliography}

\end{document}

\title{
Supplement: Suppression of bacterial rheotaxis in wavy channels
}

\author{Winfried Schmidt}
\affiliation{Theoretische Physik, Universit\"at Bayreuth, 95440 Bayreuth, Germany }
\affiliation{Universit\'e Grenoble Alpes, CNRS, LIPhy, F-38000 Grenoble, France}

\author{Igor S.\ Aranson}
\affiliation{
	Departments of Biomedical Engineering, Chemistry, and Mathematics, Pennsylvania State University,
	University Park, PA 16802, USA
}

\author{Walter Zimmermann}
\affiliation{Theoretische Physik, Universit\"at Bayreuth, 95440 Bayreuth, Germany }

\maketitle

\onecolumngrid
\setlength{\parindent}{0cm}
\setlength{\parskip}{8pt}

\subsection{Full equations of motion}
The mobility matrix in the first line of Eqs.\ (1) of the main text mediates the hydrodynamic interaction for the translational degrees of freedom and is given by the Rotne-Prager tensor for differently sized beads \cite{zuk_wajnryb_mizerski_szymczak_2014}
\begin{align}\label{eq_mm_tt}
	{\boldsymbol \mu}_{ij}^\text{tt} = \begin{cases}
		\frac{1}{8 \pi \eta \tilde{r}} \left[ \left( 1 + \frac{a_i^2 + a_j^2}{3 \tilde{r}^2} \right) {\bf 1} + \left( 1 - \frac{a_i^2 + a_j^2}{\tilde{r}^2} \right) \frac{\tilde{{\bf r}} \otimes \tilde{{\bf r}}}{\tilde{r}^2}  \right] &\text{for} \quad \tilde{r} > a_i + a_j \\
		\frac{1}{6 \pi \eta a_i a_j} \left[ \frac{16 \tilde{r}^3 ( a_i + a_j) - \left( ( a_i - a_j)^2 + 3 \tilde{r}^2 \right)^2}{32 \tilde{r}^3} {\bf 1} + 
		\frac{3 \left( (a_i - a_j)^2 - \tilde{r}^2 \right)^2}{32 \tilde{r}^3} \frac{\tilde{{\bf r}} \otimes \tilde{{\bf r}}}{\tilde{r}^2} \right] &\text{for} \quad a_i + a_j \geq \tilde{r} > a_> - a_< \\	
		\frac{1}{6 \pi \eta a_>}  {\bf 1} &\text{for} \quad \tilde{r} \leq a_> - a_<~,
	\end{cases}
\end{align}
with the bead-to-bead vector $\tilde{{\bf r}} := {\bf r}_i - {\bf r}_j$ and its modulus $\tilde{r} = |\tilde{{\bf r}}|$. The identity matrix is ${\bf 1}$ and $\tilde{{\bf r}} \otimes \tilde{{\bf r}}$ denotes the outer product of $\tilde{{\bf r}}$ with itself. In the pair $i,j$, $a_>$ and $a_<$ is the bead radius of the larger or smaller bead, respectively. For hydrodynamic interactions between two swimmer beads, this distinction is not necessary since our swimmer consists of $N$ equally sized beads with radius $a$. It is however needed for interactions of the counter-force point with the swimmer beads. The counter-force point is essentially treated as an additional bead with radius $0$,  located at ${\bf r}_\text{p}  = {\bf r}_N - 2 a \hat{\bf{e}}_s$ and subjected only to the force ${\bf F}_\text{p} = -{\bf F}_0$. The flow disturbance caused by ${\bf F}_\text{p}$ accounts for the $N+1$-st contribution to the translational and angular velocities in the sums of Eqs.\ (1) of the main text, with ${\bf F}_{N+1} = {\bf F}_\text{p}$.
Together, the forces on each point of the swimmer are
\begin{align}\label{eq_swimmer_force_free}
	{\bf F}_i = \begin{cases}
	- \nabla_i \left( V_i^\text{H} + V_i^\text{B} \right) & \text{for} \quad i \in [1,N-1] \\
	- \nabla_i \left( V_i^\text{H} + V_i^\text{B} \right) + {\bf F}_0 & \text{for} \quad i =N \\
	- {\bf F}_0 & \text{for} \quad i =N+1 
	\end{cases},
\end{align}
with $V_i^\text{H}$ and $V_i^\text{B}$ for each bead as described in the main text.
The elastic (non-active) bead forces compensate each other,
\begin{equation}
	\sum_{i=1}^{N} - \nabla_i \left( V_i^\text{H} + V_i^\text{B} \right) = {\bf 0}.
\end{equation}
From Eq. (\ref{eq_swimmer_force_free}) then follows that the total swimmer, i.e., all $N$ beads and the counter-force point, is force free,
\begin{equation}
	\sum_{i=1}^{N+1} {\bf F}_i = {\bf 0}.
\end{equation}

We note that by including only translational degrees of freedom in the equations of motion for each bead, the flow-vorticity-induced rotation of our swimmer in a shear flow with straight streamlines (e.g., a linear shear or plane Poiseuille flow) would come to a halt as soon as the swimmer axis is aligned with the flow direction. However, any object with a finite aspect ratio will perform a continuous rotation under shear flow, known as Jeffery orbits \cite{doi:10.1098/rspa.1922.0078}. For this reason, we include also rotational degrees of freedom in our model. The corresponding mobility matrices are given by \cite{zuk_wajnryb_mizerski_szymczak_2014}
\begin{align}\label{eq_mm_rr}
{\boldsymbol \mu}_{ij}^\text{rr} = \begin{cases}
- \frac{1}{16 \pi \eta \tilde{r}^3} \left( {\bf 1} - 3 \frac{\tilde{{\bf r}} \otimes \tilde{{\bf r}}}{\tilde{r}^2} \right)  &\text{for} \quad \tilde{r} > a_i + a_j \\
\frac{1}{8 \pi \eta a_i^3 a_j^3} \left( \alpha {\bf 1} + \beta \frac{\tilde{{\bf r}} \otimes \tilde{{\bf r}}}{\tilde{r}^2} \right) &\text{for} \quad a_i + a_j \geq \tilde{r} > a_> - a_< \\
\frac{1}{8 \pi \eta a_>^3} {\bf 1} &\text{for} \quad \tilde{r} \leq a_> - a_<~,
\end{cases}
\end{align}
with coefficients
\begin{align}
\notag
	\alpha &= \frac{5 \tilde{r}^6 - 27 \tilde{r}^4 \left( a_i^2 + a_j^2 \right) + 32 \tilde{r}^3 \left( a_i^3 + a_j^3 \right) - 9 \tilde{r}^2 \left( a_i^2 - a_j^2 \right)^2 - \left( a_i - a_j \right)^4 \left( a_i^2 + 4 a_i a_j + a_j^2 \right)}{64 \tilde{r}^3} \quad \text{and} \\
	\beta &= \frac{3 \left( (a_i - a_j)^2 - \tilde{r}^2 \right)^2 \left( a_i^2 + 4 a_i a_j + a_j^2 -\tilde{r}^2 \right)}{64\tilde{r}^3}
\end{align}
as well as 
\begin{align}\label{eq_mm_rt}
{\boldsymbol \mu}_{ij}^\text{rt} = \begin{cases}
\frac{1}{8 \pi \eta \tilde{r}^2} {\bf \varepsilon} \frac{\tilde{{\bf r}}}{\tilde{r}} &\text{for} \quad \tilde{r} > a_i + a_j \\
\frac{1}{16 \pi \eta a_i^3 a_j} \frac{\left( a_i - a_j + \tilde{r} \right)^2 \left( a_j^2 + 2 a_j (a_i + \tilde{r}) - 3 (a_i - \tilde{r})^2 \right)}{8 \tilde{r}^2} {\bf \varepsilon} \frac{\tilde{{\bf r}}}{\tilde{r}} &\text{for} \quad a_i + a_j \geq \tilde{r} > a_> - a_< \\
\theta (a_i - a_j) \frac{\tilde{r}}{8 \pi \eta a_i^3} {\bf \varepsilon} \frac{\tilde{{\bf r}}}{\tilde{r}} &\text{for} \quad \tilde{r} \leq a_> - a_<~,
\end{cases}
\end{align}
with $\left( {\bf {\bf \varepsilon}} \tilde{{\bf r}} \right)_{\alpha \beta} = \varepsilon_{\alpha \beta \gamma} \tilde{r}_\gamma$ and the Heaviside function $\theta (x)$. ${\boldsymbol \mu}_{ij}^\text{tr}$ is obtained by interchanging bead $i$ and $j$ in Eq.\ (\ref{eq_mm_rt}).

\begin{figure}[htb]
	\begin{center}
		\includegraphics[width=0.5\columnwidth]{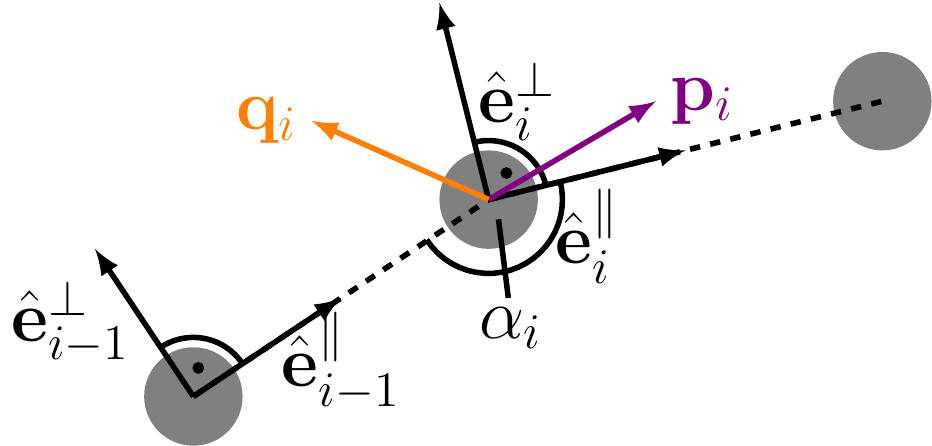} \vspace{-.5cm}
	\end{center}
	\caption{Segment of the chain around bead $i$ with opening angle $\alpha_i$. The torque according to Eq.\ (\ref{eq_torque}) penalizes deviations of the orientation vectors ${\bf p}_i$ (violet) and ${\bf q}_i$ (orange) from the local unit vectors along and perpendicular to the body axis (black dashed line).}
	\label{fig_SI_torques}
\end{figure}
We furthermore introduce two vectors ${\bf p}_i$ and ${\bf q}_i$ for each bead, characterizing its instantaneous orientation. We control the otherwise free rotation of the beads by the bead torque
\begin{align}\label{eq_torque}
{\bf T}_i = \frac{\kappa_\text{t}}{2} \left[ {\bf p}_i \times \left( \hat{\bf{e}}_i^\parallel + \hat{\bf{e}}_{i-1}^\parallel \right) + {\bf q}_i \times \left( \hat{\bf{e}}_i^\perp + \hat{\bf{e}}_{i-1}^\perp \right) \right].
\end{align}
Here, $\kappa_\text{t}$ is the torque strength, $\hat{\bf{e}}_i^\parallel$ the unit vector pointing from bead $i$ in the direction of the next bead in the chain. Furthermore, $\hat{\bf{e}}_i^\perp$ is the unit vector perpendicular to $\hat{\bf{e}}_i^\parallel$ which is obtained by shifting the polar angle of $\hat{\bf{e}}_i^\parallel$ in spherical coordinates by $\pi/2$ (cf.\ Fig.\ \ref{fig_SI_torques}). The orientation of each bead is then evolved via 
\begin{equation}
	\dot{{\bf p}}_i = {\boldsymbol \Omega}_i \times {\bf p}_i \quad \text{and} \quad \dot{{\bf q}}_i = {\boldsymbol \Omega}_i \times {\bf q}_i.
\end{equation}
In the case of a straight (undeformed) chain, the torque for bead $i$ according to Eq.\ (\ref{eq_torque}) is minimal when ${\bf p}_i$ points towards the neighboring bead to the right and ${\bf q}_i$ in the direction of $\hat{\bf{e}}_i^\perp$. Deviations of ${\bf p}_i$ or ${\bf q}_i$ from this configuration are penalized with a restoring torque ${\bf T}_i$ which couples to the velocity and angular velocity of all beads of the chain via Eqs.\ (1) in the main text. As a result, rotations of individual beads translate to a rotation of the entire chain. As we show {\color{red} below}, by this we reproduce Jeffery orbits of the stiff passive chain in a shear flow.
\subsection{Jeffery dynamics of a passive rod in linear shear flow}
\begin{figure}[htb]
	\begin{center}
		\includegraphics[width=0.5\columnwidth]{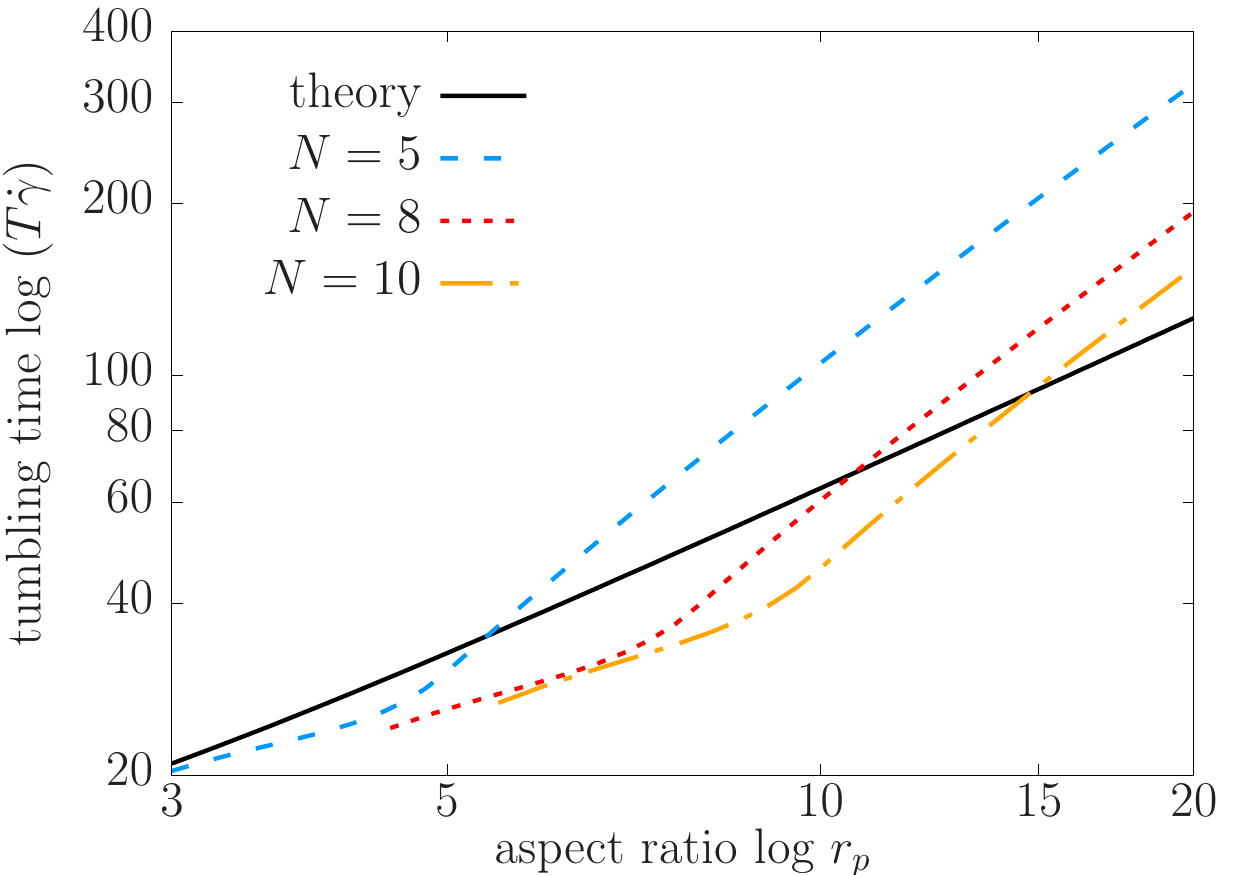} \vspace{-.5cm}
	\end{center}
	\caption{Tumbling time $T$ of a passive, stiff chain of beads in a linear shear flow with shear rate $\dot{\gamma}$ as function of the aspect ratio $r_\text{p}$ for different numbers of beads $N$. Blue dashed lines correspond to $N=5$, red dashed lines to $N=8$, and orange dashed lines to $N=10$. The black bold line shows the theoretical prediction according to Eq.\ (\ref{eq_time_jeffery_theory}). Larger values of $N$ match the prediction when higher aspect ratios are chosen (intersections with the black line).}
	\label{fig_jeffery_orbits_period}
\end{figure}

In a linear shear flow, the time period for a rigid rod with aspect ratio $r_\text{p}$ to perform one Jeffery orbit has been derived as \cite{doi:10.1122/1.550604}
\begin{equation}\label{eq_time_jeffery_theory}
T = \frac{2 \pi}{\dot{\gamma}} \left( r_\text{p} + \frac{1}{r_\text{p}}  \right)~,
\end{equation}
which is also referred to as the tumbling time. We use Eq.\ (\ref{eq_time_jeffery_theory}) to fit our parameters in order to obtain a realistic behavior of our model. For this we focus on a passive chain of beads ($F_0 = 0$) with negligible deformability ($k= \kappa = 100$) in linear shear flow ${\bf u} ({\bf r}_i) = \dot{\gamma} y_i \hat{\bf{e}}_x$ with shear rate $\dot{\gamma} = 0.1$. We choose $a=0.5$ as a fixed parameter and vary $b$ to adapt the aspect ratio $r_\text{p} = 1+(N-1)b/(2a)$. 
Fig.\ \ref{fig_jeffery_orbits_period} shows the simulation results for three rods with different numbers of beads $N$ together with the theoretical prediction of Eq.\ (\ref{eq_time_jeffery_theory}). The tumbling time of our model lies below the theoretical prediction for small aspect ratios and grows according to a power law $T \propto r_\text{p}^\delta$ for large $r_\text{p}$, where $\delta > 1$ seems to be independent of $N$. For each number of beads, the numerically determined values for $T$ match with Eq.\ (\ref{eq_time_jeffery_theory}) for one aspect ratio. In this work we limit our studies to $N=5$, where we find optimal agreement for an aspect ratio of $5.38$, corresponding to $b=1.095$. These parameters are used for all of our simulations.
\subsection{Intrinsic swimming speed}
\begin{figure}[htb]
	\begin{center}
		\includegraphics[width=0.5\columnwidth]{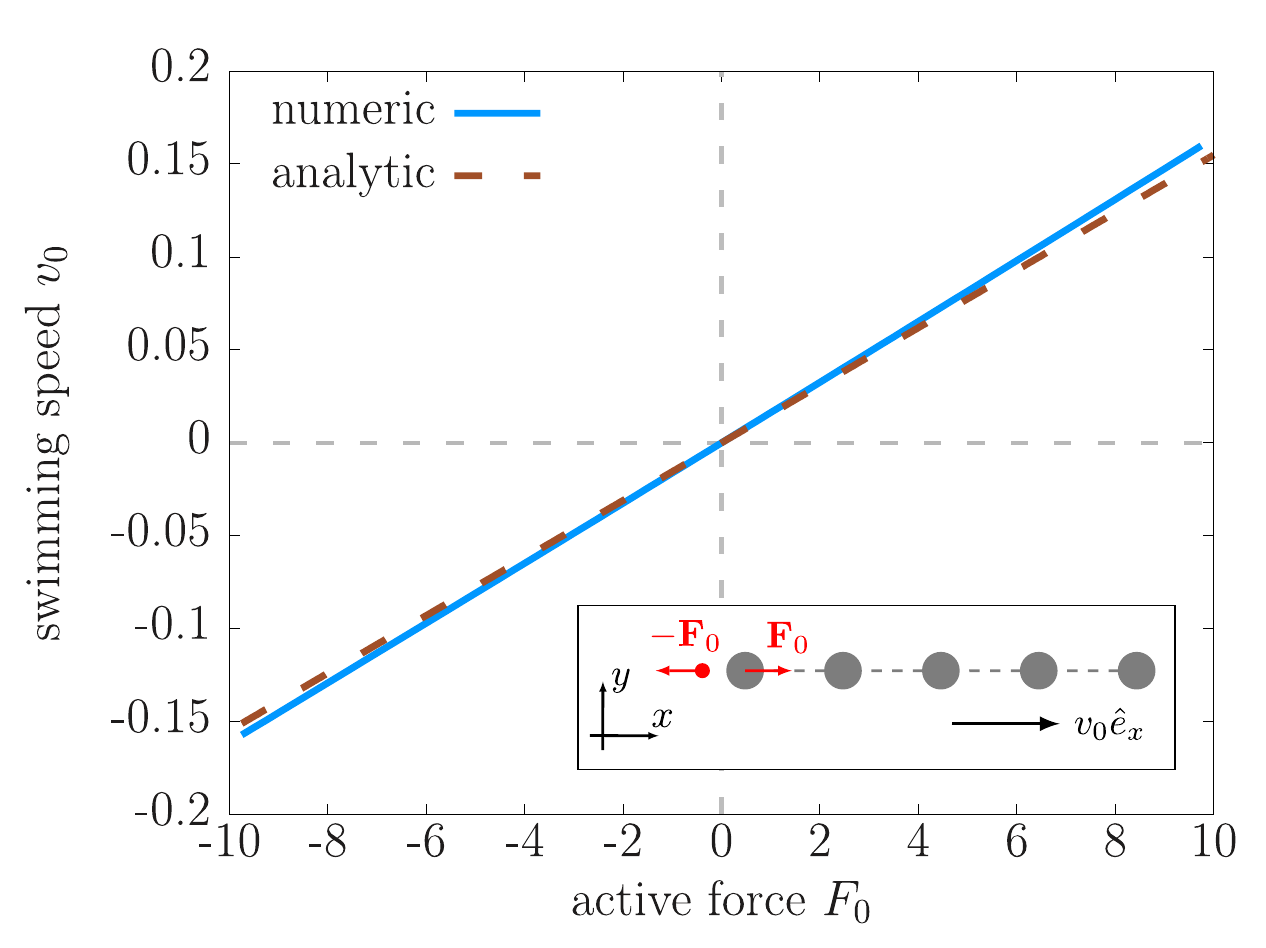} \vspace{-.5cm}
	\end{center}
	\caption{Calibration curve for the intrinsic swimming speed $v_0$ (propulsion speed without incident flow) as function of the active force's absolute value $F_0$ for the stiff swimmer. Positive $F_0$ correspond to a pusher, negative $F_0$ to a puller. The results of the analytical calculation (brown dashed line) according to Eq.\ (\ref{eq_app_analytic_result}) and the numerical simulation (blue bold line) is shown. Inset: Sketch of the swimmer with $N=5$ beads. The force dipole is located at its left, the resulting swimming velocity $v_0 \hat{\bf{e}}_x$ is directed along the $x$-axis.}
	\label{fig_calibration_curve_intrisic_swim_speed}
\end{figure}

In our swimmer model, self-propulsion is achieved by the active force ${\bf F}_0$ at the swimmer's rear end. The relevant physical quantity, however, is the resulting intrinsic swimming speed $v_0$. In a quiescent fluid ($u_0 = 0$) we approximate $v_0$ as function of $F_0$ analytically for a non-deformable swimmer. For this, we neglect bead rotations as well as the hydrodynamic flow field created by the elastic forces and take into account only the flow field originating from the pair of active forces.
We consider a swimmer initially aligned with the $x$-axis with beads positioned at ${\bf r}_i = (- (i - 1) b, 0, 0)$ with $i = 1, ..., N$, and the counter-force point located at ${\color{red}{\bf r}_\text{p} = } (- (N-1) {\color{red}b} - 2a, 0, 0)$ (cf.\ inset in Fig.\ \ref{fig_calibration_curve_intrisic_swim_speed}). 
We consider a stiff swimmer, so the motion of the beads relatively to each other can be neglected. With Eqs.\ (1) of the main text, the simplified equations of motion for the swimmer velocity then yield
\begin{align}\label{eq_app_eq_of_motion_calibration_curve}
	\dot{{\bf r}}_\text{c} = \frac{1}{N} \sum_{i = 1}^{N} \dot{{\bf r}}_i = \frac{1}{N} \sum_{i = 1}^{N} \left( \frac{{\bf F}_i}{6 \pi \eta a} + \sum_{j = 1, j \neq i}^{N} {\boldsymbol \mu}_{i j}^\text{tt} {\bf F}_j + {\boldsymbol \mu}_{i \text{p}}^\text{tt} {\bf F}_\text{p} \right),
\end{align}
where ${\boldsymbol \mu}_{i \text{p}}^\text{tt}$ describes the hydrodynamic interaction between bead $i$ and the counter-force point, according to Eq.\ (\ref{eq_mm_tt}).
Using ${\bf F}_N = F_0 \hat{\bf{e}}_x$, ${\bf F}_\text{p} = - F_0 \hat{\bf{e}}_x$ and ${\bf F}_i = {\bf 0}$ for $i = 1,...,N-1$, it follows from Eq.\ (\ref{eq_mm_tt}) $\dot{{\bf r}}_\text{c} = v_0 \hat{\bf{e}}_x$ with
\begin{align}\label{eq_app_analytic_result}
	v_0 = \frac{F_0 }{4 \pi \eta N}\Bigg\{\ \frac{5}{24 a}  
	+ \sum_{i = 1}^{N - 1} \frac{1}{(N - i) b} \left[ 1 - \frac{2 a^2}{3 (N-i)^2 b^2} \right]  
	- \frac{1}{\left( (N - i) b + 2a \right)} \left[ 1 - \frac{a^2}{3 \left( (N-i) b + 2a \right)^2} \right]  \Bigg\}\ .
\end{align}
This result is shown in Fig.\ \ref{fig_calibration_curve_intrisic_swim_speed} for $N=5$ together with the numerically obtained values for the intrinsic swimming speed. In the simulation we choose $k= \kappa = 100$ to ensure negligible deformation. For $F_0 > 0$ a pusher-type flow field is created by the two opposite active forces and the swimmer moves in positive $x$-direction, whereas for $F_0 < 0$ we obtain a puller-type flow field and motion in negative $x$-direction. Differences between the analytical and numerical results arise from the approximation of zero bead forces, but overall good agreement is observed.
We use the numerically obtained values for $v_0$ as function of $F_0$ in the main text.
\subsection{Swimmer behavior in plane Poiseuille flow}
\begin{figure}[htb]
	\begin{center}
		\includegraphics[width=0.5\columnwidth]{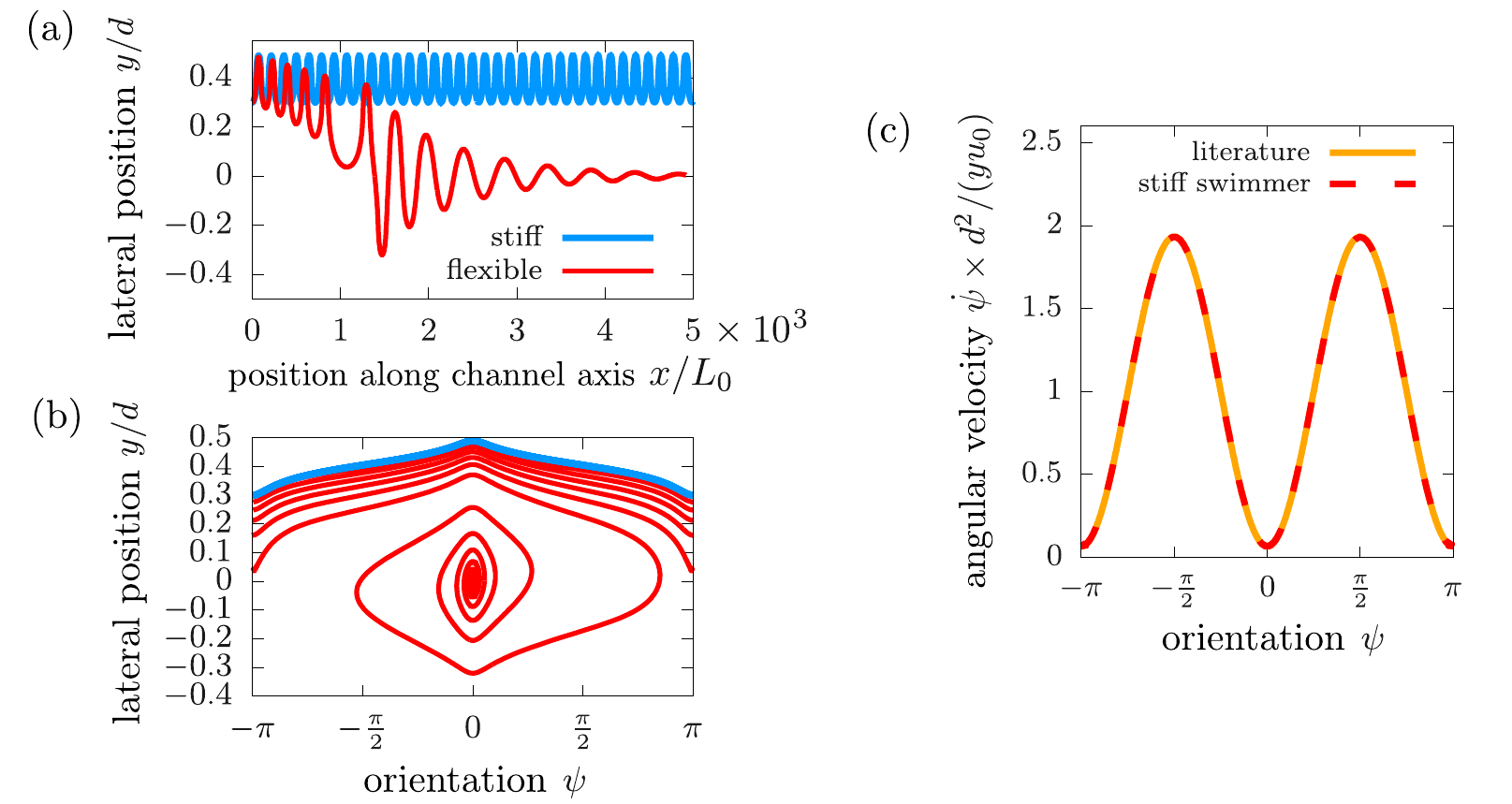} \vspace{-.5cm}
	\end{center}
	\caption{
	(a) 	Real space  and (b) corresponding phase space  trajectory of a flexible (red line) and a stiff swimmer (blue line) in plane Poiseuille flow with initial position $y_{\text{c}, 0} = 0.3 d$ and orientation $\psi_0 = \pm \pi$ (downstream swimming). $L_0$ is the initial swimmer length and $d$ the channel half-width. A stiff swimmer performs a tumbling motion around a fixed off-centered position (periodic phase space orbits). A flexible swimmer tumbles and simultaneously migrates towards the channel center, eventually switching to a swinging motion with decaying amplitude. (c) Angular velocity $\dot{\psi}$ of a stiff swimmer (red dashed line) vs its instantaneous orientation $\psi$. The result agrees well with the theoretical prediction from the literature [Eq.\ (\ref{eq_angular_velocity_lit}), orange bold line].
	}
	\label{fig_swinging_tumbling} 
\end{figure}

We further validate our model for the swimmer by analyzing its behavior in a plane Poiseuille flow, corresponding to $\varepsilon = 0$ in Eq.\ (3) of the main text. For this we employ stiff ($\kappa=100$) and flexible swimmers ($\kappa=0.5$) and an activity of $F_0 = 0.3$ with resulting intrinsic swimming speed of $v_0 = 4.872 \times 10^{-3}$.
Fig.\ \ref{fig_swinging_tumbling}(a) shows the trajectory of a stiff swimmer (blue line) with negligible deformation in the flow. As reported previously \cite{PhysRevLett.108.218104, zottl2013periodic}, such a swimmer either performs a swinging motion or tumbles around a constant off-centered position in the channel that depends on its initial conditions. The corresponding periodic phase space orbit is shown in Fig.\ \ref{fig_swinging_tumbling}(b). For an elongated swimmer as employed by us, the angular velocity becomes a function of the instantaneous orientation angle, namely \cite{zottl2013periodic}
\begin{equation}\label{eq_angular_velocity_lit}
	\dot{\psi} = \frac{u_0}{d^2} y \left(1 - G \cos (2 \psi)\right).
\end{equation}
For a stiff swimmer we find very good agreement with Eq.\ (\ref{eq_angular_velocity_lit}), as Fig.\ \ref{fig_swinging_tumbling}(c) shows.
By contrast, a flexible swimmer [real space trajectory in Fig.\ \ref{fig_swinging_tumbling}(a), phase space trajectory in Fig.\ \ref{fig_swinging_tumbling}(b), red lines] in addition to its tumbling motion shows a lateral drift towards the channel center, enabling a tumbling swimmer to switch to swinging. This transition occurs below a critical $y$-position when the flow vorticity $\nabla \times {\bf u} = 2 u_0 y /d^2 \hat{\bf{e}}_z$ is not sufficiently strong anymore to reorient the swimmer before it crosses the center [at $x/L_0 \approx 1.39 \times 10^{3}$ in Fig.\ \ref{fig_swinging_tumbling}(a)]. The amplitude of the swinging motion subsequently decreases, and the swimmer  approaches an attractor of upstream swimming at the centerline [$(y_\text{c},\psi) = (0,0)$]. This  behavior has been found previously for microswimmers with flexible flagella \cite{tournus2015flexibility, C9SM00717B}.
\subsection{Analytical derivation of the flow field in the wavy channel}
\label{sec_app_analytical}
Here we derive the analytical solution for the flow field through serpentine wavy channels. The calculations are similar to the considerations in Refs.\ \cite{tsangaris1984laminar, tsangaris1986laminar, PhysRevLett.122.128002}. We have to solve
\begin{equation}\label{eq_basic_equation}
\nabla^{4} \Psi(x, y) = 0,
\end{equation}
where $\Psi(x, y)$ is the stream function and $\nabla^{4} = \left( \partial^2_x + \partial^2_y \right)^2$ the biharmonic operator. The stream function fulfills the conditions
\begin{align}
u_{x} (x, y) = \partial_y \Psi (x, y), \qquad u_{y} (x, y) = - \partial_x \Psi (x, y).
\end{align}
The walls of the wavy channel are at
\begin{equation}
y_\text{w} (x) = d \left[\pm 1 + \varepsilon \sin \left( \frac{2 \pi x}{\lambda} \right) \right].
\end{equation}
One has the no-slip boundary conditions
\begin{align}\label{eq_noslip_physical_coords}
u_{x} \left( x, y = y_\text{w} (x) \right) = 0, \qquad u_{y} \left( x, y = y_\text{w} (x) \right) = 0.
\end{align}
Furthermore we assume point symmetry of the flow field around the origin of coordinates which is given by
\begin{align}\label{eq_phys_point_sym}
u_{x} \left( - x, - y \right) = u_{x} \left(x, y \right), \qquad u_{y} \left( - x, - y \right) = u_{y} \left(x, y \right).
\end{align}
We start by introducing the dimensionless coordinates
\begin{equation}\label{eq_dimensionless_variables}
x^\prime := \frac{x}{\lambda}, \qquad y^\prime:= \frac{y}{d}, \qquad \Psi^\prime := \frac{\Psi}{u_0 d},
\end{equation}
where $u_0$ is the characteristic flow speed. With this, the plane coordinates
\begin{align}
\notag
\eta (x^\prime) &:= x^\prime \\
\zeta (x^\prime,y^\prime) &:= y^\prime - \varepsilon \sin \left( 2 \pi x^\prime \right)
\end{align}
can be introduced where the walls are located at $\zeta = \pm 1$. The boundary conditions according to Eq.\ (\ref{eq_noslip_physical_coords}) then transform into
\begin{equation}\label{eq_BCs_new_cos}
\left. \partial_\eta \Psi  (\eta, \zeta) \right\vert_{\zeta = \pm 1} = 0, \qquad \left. \partial_\zeta \Psi (\eta, \zeta) \right\vert_{\zeta = \pm 1} = 0,
\end{equation}
where $\Psi (\eta, \zeta)$ is the stream function in plane coordinates which we have to calculate. For this, we translate the point symmetry according to Eq.\ (\ref{eq_phys_point_sym}) into the new coordinates which yields
\begin{align} \label{eq_point_sym_new}
\notag
\partial_\zeta \Psi \left(- \eta, - \zeta \right) &= \partial_\zeta \Psi \left(\eta, \zeta \right),  \\
- \partial_\eta \Psi \left(- \eta, - \zeta \right) &= - \partial_\eta \Psi \left(\eta, \zeta \right).
\end{align}
This implies that $\Psi(\eta, \zeta)$ is an odd function of its arguments, namely $- \Psi(\eta, \zeta) = \Psi (-\eta, -\zeta)$. Eq.\ (\ref{eq_basic_equation}) then transforms into
\begin{equation}\label{eq_basic_equation_new_cos}
\bar{\Delta} \bar{\Delta} \Psi(\eta, \zeta) = 0,
\end{equation}
where $\bar{\Delta}$ is the Laplace operator in the plane coordinates.
Since $\varepsilon$ is assumed to be small, a perturbation analysis is possible and thus the solution of Eq.\ (\ref{eq_basic_equation_new_cos}) can be written as
\begin{equation}\label{eq_taylor_psi}
\Psi \approx \Psi_0 + \varepsilon \Psi_1,
\end{equation}
where $\Psi_0$ and $\Psi_1$ have to obey the boundary conditions (\ref{eq_BCs_new_cos}) separately. Substituting Eq.\ (\ref{eq_taylor_psi}) into Eq.\ (\ref{eq_basic_equation_new_cos}) and sorting the resulting terms with respect to orders in $\varepsilon$ yields the equation for the zeroth order,
\begin{equation}
\partial^4_\zeta \Psi_0 (\zeta) =0,
\end{equation}
which has the solution
\begin{equation}\label{eq_psi0_sol}
\Psi_0 (\zeta) = \zeta - \frac{\zeta^3}{3}.
\end{equation}
This can be substituted into the equation for the first order in $\varepsilon$, resulting in
\begin{align}\label{eq_psi1_first_order}
\partial_\zeta^4 \Psi_1 \left( \eta,\zeta \right) 
+ 2 \partial_\zeta^2 \partial_\eta^2 \Psi_1 \left( \eta,\zeta \right) 
+ \partial_\eta^4 \Psi_1 \left( \eta,\zeta \right)
	+ \left(  \left( 
16\,{\zeta}^{2}-16 \right) {\pi}^{4}-16\,{\pi}^{2} \right) \sin
\left( 2 \pi \eta \right) =0.
\end{align}
The solution of Eq.\ (\ref{eq_psi1_first_order}) which is obtained by separation of variables is
\begin{align}
\Psi_1 (\eta, \zeta)= \frac{\sin(2 \pi \eta)}{4 \pi + \sinh (4 \pi)} \biggl[ -4 \sinh(2 \pi) \cosh(2 \pi \zeta) + 4 \zeta \sinh(2 \pi \zeta) \cosh(2 \pi) - 4 \left( \zeta^2 - 1 \right) \left( \frac{1}{2} \cosh(2 \pi) \sinh(2 \pi) + \pi \right) \biggr],
\end{align}
and thus the total solution for the stream function up to $\mathcal{O}(\varepsilon^1)$ is
\begin{align}\label{eq_psi_eta_zeta_full}
\notag \Psi (\eta, \zeta) = \ & \zeta - \frac{\zeta^3}{3} + \varepsilon 
\frac{4 \sin (2 \pi \eta)}{4 \pi + \sinh (4 \pi)} \biggl[ - \sinh(2 \pi) \cosh(2 \pi \zeta)+ \zeta \sinh(2 \pi \zeta) \cosh(2 \pi)  \\
& - \left( \zeta^2 - 1 \right) \left( \frac{1}{2} \cosh(2 \pi) \sinh(2 \pi) + \pi \right) \biggr].
\end{align}
The expressions for the flow field in the physical coordinates can be calculated from Eq.\ (\ref{eq_psi_eta_zeta_full}) according to
\begin{align}\label{eq_flow_field_result_physical}
\notag
u_x^\prime (x^\prime,y^\prime) &= \partial_\eta \Psi(\eta, \zeta) \frac{d}{d y^\prime} \eta (x^\prime,y^\prime) + \partial_\zeta \Psi(\eta, \zeta) \frac{d}{d y^\prime} \zeta (x^\prime,y^\prime), \\
u_y^\prime (x^\prime,y^\prime) &= - \partial_\eta \Psi(\eta, \zeta) \frac{d}{d x^\prime} \eta (x^\prime,y^\prime) - \partial_\zeta \Psi(\eta, \zeta) \frac{d}{d x^\prime} \zeta (x^\prime,y^\prime)
\end{align}
By this we obtain solutions for $u_x^\prime (x^\prime,y^\prime)$ and $u_y^\prime (x^\prime,y^\prime)$ which contain orders in $\varepsilon$ larger than one. Therefore, one has to expand the solutions again with respect to $\varepsilon$ and truncate the result after the first order in $\varepsilon$. The final expressions for the flow field are
\begin{align}\label{eq_wavy_flow_result}
\notag 
u_{x} (x, y) &= u_0 \bigg[ 1 - \frac{y^{2}}{d^2} + \varepsilon \frac{\sin (\frac{2 \pi x}{\lambda})}{4 \pi + \sinh \left( 4 \pi \right)} \bigg( \left( -8 \pi \sinh(2 \pi) + 4 \cosh(2 \pi) \right) \sinh \left( \frac{2 \pi y}{d} \right)  \\
\notag
&+ \left( 8 \pi \cosh(2 \pi) \cosh \left( \frac{2 \pi y}{d} \right) - 4 \cosh(2 \pi) \sinh(2 \pi) + 2 \sinh(4 \pi) \right) \frac{y}{d}  \bigg) \bigg], \\
\notag u_{y} (x, y) &= - u_0 \ \varepsilon \cos \left( \frac{2 \pi x}{\lambda} \right) \frac{2 \pi}{4 \pi + \sinh \left( 4 \pi \right)} \Bigg[4 \cosh(2 \pi) \frac{y}{d} \sinh \left( \frac{2 \pi y}{d} \right) \\
&-4 \sinh(2 \pi) \cosh \left( \frac{2 \pi y}{d} \right) + \left( \sinh(4 \pi) - 2 \cosh(2 \pi) \sinh(2 \pi) \right) \frac{y^{2}}{d^2}
\Bigg].
\end{align}
With the functions
\begin{align}
	\notag U_1 (x,y) &:= \frac{\sin (\frac{2 \pi x}{\lambda})}{4 \pi + \sinh \left( 4 \pi \right)} \bigg( \left( -8 \pi \sinh(2 \pi) + 4 \cosh(2 \pi) \right) \sinh \left( \frac{2 \pi y}{d} \right) \\
	\notag
	&+ \left( 8 \pi \cosh(2 \pi) \cosh \left( \frac{2 \pi y}{d} \right) - 4 \cosh(2 \pi) \sinh(2 \pi) + 2 \sinh(4 \pi) \right) \frac{y}{d}  \bigg), \\
	\notag U_2 (x,y) &:= -  \cos \left( \frac{2 \pi x}{\lambda} \right) \frac{2 \pi}{4 \pi + \sinh \left( 4 \pi \right)} \Bigg[4 \cosh(2 \pi) \frac{y}{d} \sinh \left( \frac{2 \pi y}{d} \right) \\
	&-4 \sinh(2 \pi) \cosh \left( \frac{2 \pi y}{d} \right) + \left( \sinh(4 \pi) - 2 \cosh(2 \pi) \sinh(2 \pi) \right) \frac{y^{2}}{d^2}
	\Bigg]
\end{align}
the flow field can be written as in Eq.\ (3) of the main text.
In the special case of no wall modulation ($\varepsilon=0$), we recover the plane Poiseuille flow profile
\begin{align}
\notag
u_{x} &= u_0 \left( 1 - \frac{y^{2}}{d^2} \right),\\
u_{y} &= 0
\end{align}
from Eqs.\ (\ref{eq_wavy_flow_result}).

\newpage
\subsection{Influence of swimmer activity on the resonance curve}
\begin{figure}[htb]
	\begin{center}
		\includegraphics[width=0.7\columnwidth]{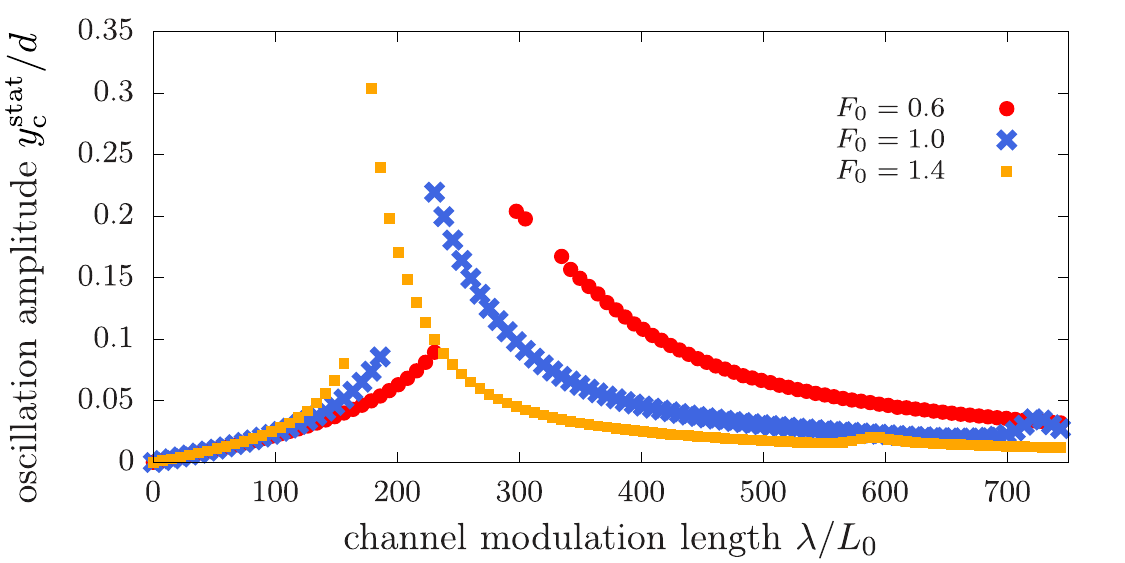} \vspace{-.5cm}
	\end{center}
	\caption{
		Long-time oscillation amplitude $y_\text{c}^\text{stat}$ in units of the channel half-width $d$ as function of the channel modulation length $\lambda$ in units of the swimmer's initial length $L_0$. The results for three different activities $F_0 = 0.6$ (red circles), $F_0 = 1.0$ (blue crosses) and $F_0 = 1.4$ (orange squares) are shown. 
	}
	\label{fig_resonance_activity} 
\end{figure}

Fig.\ \ref{fig_resonance_activity} shows three resonance curves for different swimmer activities. Generally, smaller activities result in larger oscillation amplitudes. Furthermore, we observe a shift of the maximum towards larger values of $\lambda$ for decreasing $F_0$. This is in accordance with Eq.\ (4) in the main text which predicts the resonance wavelength to decrease as function of $v_0$ in the regime of downstream drift. 
%In addition to the global maximum a smaller local maximum is observed in the curves for $F_0 = 1.0$ and  $F_0 = 1.4$ for large channel wavelengths. 
%
%
%\wsnote{
%
\subsection{Fourier spectra of the swimmer trajectory}
\begin{figure}[htb]
	\begin{center}
		\includegraphics[width=0.7\columnwidth]{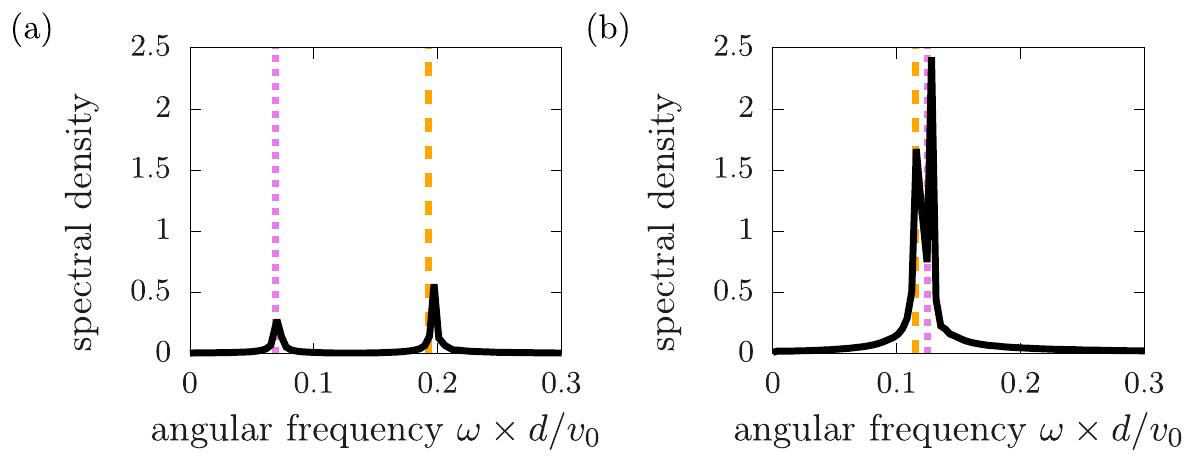} \vspace{-.5cm}
	\end{center}
	\caption{
		Spectra $P (\omega)$ of the swimmer's lateral position as function of time, $y_\text{c}(t)$, for  $u_0 / v_0 = 0.554$ [(a), off-resonant] and $u_0 / v_0 = 0.197$ [(b), close to resonance], and the predictions for $\omega_0$ (orange dashed lines) and $\omega_\text{Ch}$ (pink dotted lines).
	}
	\label{fig_fourier_spectra} 
\end{figure}

We calculate the Fourier spectrum of the swimmer's lateral position, $y_\text{c}(t)$. Fig.\ \ref{fig_fourier_spectra}(a) shows the spectrum for a swimmer with $\kappa = 3$ for $u_0 / v_0 = 0.554$ which leads to an off-resonant oscillation with small amplitude (cf. Fig. 4 of the main text). We observe two maxima in the spectrum that are close to the theoretical approximations for $\omega_0$ and $\omega_\text{Ch}$, as described in the main text. For a flow strength close to the resonance case [$u_0 / v_0 = 0.197$, cf. Fig. \ref{fig_fourier_spectra}(b)], both the theoretical and numerical locations for the maxima are shifted closer towards each other. 
\subsection{Repulsive wall potential}
To prevent the swimmer from penetrating the channel wall we include a short-range repulsive potential \cite{doi:10.1063/1.1674820}
\begin{equation}
V(r_i^\text{w}) = \begin{cases}
V^*(r_i^\text{w}) \quad &\text{for} \quad r_i^\text{w} \leq r^\text{w,c} \\
0 \quad &\text{for} \quad r_i^\text{w} > r^\text{w,c} \\
\end{cases}
\end{equation}
with
\begin{equation}\label{eq_repulsive_wall_potantial2}
V^*(r_i^\text{w}) = 4 V_0 \left[ \left( \frac{\sigma}{r_i^\text{w}} \right)^{12} - \left( \frac{\sigma}{r_i^\text{w}} \right)^{6} \right].
\end{equation}
Here $r_i^\text{w}$ is the shortest distance between position of bead $i$ to the wavy channel wall, $V_0 = 0.05$ the repulsion energy and $\sigma = 2^{-\frac{1}{6}} a$ the repulsion length. The cut-off distance $r^\text{w,c} = 2^\frac{1}{6} \sigma$ is chosen in such a way that only the repulsive part of eq.\ (\ref{eq_repulsive_wall_potantial2}) is taken into account.
%

%\bibliographystyle{prsty}

%\bibliography{bib_swim.bib}